\documentclass[preprint,journal]{vgtc}       

\ifpdf
  \pdfoutput=1\relax                   
  \usepackage{graphicx}                
  \DeclareGraphicsExtensions{.pdf,.png,.jpg,.jpeg} 
\else
  \usepackage{graphicx}                
  \DeclareGraphicsExtensions{.eps}     
\fi%

\graphicspath{{figures/}{pictures/}{images/}{./}} 

\usepackage{microtype}                 
\PassOptionsToPackage{warn}{textcomp}  
\usepackage{textcomp}                  
\usepackage{mathptmx}                  
\usepackage{times}                     
\usepackage{cite}                      
\usepackage{tabu}                      
\usepackage{booktabs}                  

\usepackage{fancyvrb}
\usepackage{tgcursor}

\usepackage{tabularx}
\usepackage{makecell}

\usepackage{soul} 
\usepackage{dirtytalk} 
\usepackage{csquotes} 
\usepackage{censor} 
\usepackage[table]{xcolor}

\definecolor{yi}{rgb}{0.9,1,0.96}
\definecolor{heer}{rgb}{1,1,0.8}

\newcommand{\hlc}[2][yellow]{{\sethlcolor{#1}\hl{#2}}}
\newcommand{\hlyi}[2][black]{\textcolor{black}{\hlc[yi]{\textit{#2}}}}
\newcommand{\hlheer}[2][black]{\textcolor{black}{\hlc[heer]{\textit{#2}}}}

\onlineid{1138}

\vgtccategory{Research}
\vgtcpapertype{system}

\title{Animated Vega-Lite: Unifying Animation with a \\ Grammar of Interactive Graphics}

\author{Jonathan Zong*, Josh Pollock*, Dylan Wootton, Arvind Satyanarayan}
\authorfooter{
\item Jonathan Zong and Josh Pollock are co-first authors.
\item
 The authors are with MIT CSAIL. E-mails: \{jzong, jopo, dwootton, arvindsatya\}@mit.edu.
}

\shortauthortitle{Zong & Pollock \MakeLowercase{\textit{et al.}}: A Unified Grammar of Interactive and Animated Visualizations}

\abstract{
We present Animated Vega-Lite, a set of extensions to Vega-Lite that model animated visualizations as time-varying data queries. 
In contrast to alternate approaches for specifying animated visualizations, which prize a highly expressive design space, Animated Vega-Lite prioritizes unifying animation with the language's existing abstractions for static and interactive visualizations to enable authors to smoothly move between or combine these modalities.
Thus, to compose animation with static visualizations, we represent time as an \textit{encoding channel}. Time encodings map a data field to animation keyframes, providing a lightweight specification for animations without interaction. To compose animation and interaction, we also represent time as an \textit{event stream}; Vega-Lite selections, which provide dynamic data queries, are now driven not only by input events but by timer ticks as well.
We evaluate the expressiveness of our approach through a gallery of diverse examples that demonstrate coverage over taxonomies of both interaction and animation. 
We also critically reflect on the conceptual affordances and limitations of our contribution by interviewing five expert developers of existing animation grammars. 
These reflections highlight the key motivating role of in-the-wild examples, and identify three central tradeoffs: the language design process, the types of animated transitions supported, and how the systems model keyframes.
}


\vgtcinsertpkg

\begin{document}


\firstsection{Introduction}

\maketitle

Rapid prototyping is critical to the visualization authoring process.
When making an explanatory graphic, rapid prototyping allows a visualization author to evaluate candidate designs before committing to refining one in detail.
For exploratory data analysis, rapid prototyping is equally key as visualization is just one part of a broader workflow, with analysts focused on producing and analyzing a chart to yield insight or seed further analysis.
However, consider the friction of visualizing faceted data: an author might choose between depicting facets as a small multiples display, on-demand via interaction (e.g., dynamic query widgets), or played sequentially via animation. 
These designs make different trade-offs between time and space and, as a result, research results suggest they afford readers different levels of clarity, time commitment, and visual interest~\cite{robertson_effectiveness_2008}.
Despite these differences, the designs express a shared goal\,---\,to visualize different groupings of the data\,---\,and a visualization author might reasonably expect to be able to easily move between the three to make the most appropriate choice.

Unfortunately, existing visualization toolkits can present a highly viscous~\cite{thomas_rg_green_cognitive_1989} specification process when navigating this time-space trade-off. 
One class of toolkits supports either interaction or animation, but not both.
Such systems include Vega~\cite{satyanarayan_declarative_2014} and Vega-Lite~\cite{satyanarayan_vega-lite_2017}\,---\,which offer interaction primitives in the form of \textit{signals} and \textit{selections} but do not provide abstractions for animation\,---\,as well as gganimate~\cite{thomas_lin_pedersen_grammar_2019}, Data Animator~\cite{thompson_data_2021}, Canis / CAST~\cite{ge_canis_2020, ge_cast_2021}, and Gemini/Gemini\textsuperscript{2}~\cite{kim_gemini_2021, kim_gemini2_2021}\,---\,which express animation in terms of transitions between discrete visualization states known as \textit{keyframes} but do not provide treatment for interaction.
As a result, these systems force visualization authors to prematurely commit~\cite{thomas_rg_green_cognitive_1989} to either an interaction- or animation-friendly abstraction when choosing their prototyping tool, and thus limit authors' ability to explore alternative designs.
A second class of toolkits (including D3~\cite{bostock_d_2011} and Plotly~\cite{noauthor_plotly_2012}) support both modalities but do so via largely distinct abstractions (namely, \emph{transitions} or \emph{frames} for animation, and \emph{event handlers} or a typology of techniques for interaction). 
Thus, an author must often either restructure or rewrite their specifications to consider interaction and animation in parallel.

In this paper, we present Animated Vega-Lite: extensions to Vega-Lite to support data-driven animation.
Its design is motivated by the key insight that interaction and animation are parallel concepts (\autoref{sec:vlia-by-example}). Whereas interactions transform data (e.g. filtering) and update visual properties (e.g. re-coloring marks) in response to \textit{user} input, animations do the same in response to a \emph{timer}. 
From this perspective, interactive and animated visualization techniques occupy a spectrum of dynamic, event-driven behaviors.
Thus, with Animated Vega-Lite, animated visualizations (like their interactive counterparts) are modeled as time-varying data queries\,---\,an approach that allows us to provide a \textit{unified} set of abstractions for static, interactive, and animated visualizations. 


Animated Vega-Lite offers two abstractions of time that allow animations to compose with Vega-Lite's existing grammars of static and interactive visualizations (\autoref{sec:grammar}). 
From the perspective of interaction, time is an \textit{event stream}: a source of events analogous to \verb|click|s and \verb|keypress|es produced by a user.
These events drive Vega-Lite \textit{selections}, which apply dynamic data queries to visual encodings. 
Thus, by modeling time as an event stream, users can seamlessly specify and move between interactive and animated behavior in the same specification. 
From the perspective of Vega-Lite's grammar of graphics, time is an \textit{encoding channel}. 
Just as \texttt{x} and \texttt{y} encodings map data values to spatial positions measured in pixels, a \texttt{time} encoding maps data values to temporal positions measured in elapsed milliseconds. 
Compared to the event stream abstraction, the encoding channel abstraction is lighter-weight, but less expressive. 
This allows a visualization author to get started quickly with an animated chart and to move easily between an animated and a faceted visualization by switching a \verb|time| channel for a \verb|row| or \verb|column| one.
And, for added customizability, users can always turn a time-as-encoding specification into a time-as-event-stream one. 







We implement a prototype compiler that synthesizes a low-level Vega specification with shared reactive logic for interaction and animation (\autoref{sec:implementation}).
Following best practices~\cite{ren_reflecting_2018}, we assess our contribution with multiple evaluation methods.
Through a diverse example gallery (\autoref{sec:example-gallery}), we demonstrate that Animated Vega-Lite covers much of Yi et al.'s interaction taxonomy~\cite{yi_toward_2007} and Heer \& Robertson's animation taxonomy~\cite{heer_animated_2007} while preserving Vega-Lite's low viscosity and systematic generativity.
We also interview five expert developers of four existing animated visualization grammars~\cite{thomas_lin_pedersen_gganimate_2019, thompson_data_2021, ge_canis_2020, ge_cast_2021, kim_gemini_2021, kim_gemini2_2021} to critically reflect~\cite{satyanarayan_critical_2019} on the tradeoffs, conceptual affordances, and limitations of our system (\autoref{sec:eval}).
We discuss the important role example visualizations play in grammar design and analyze three areas of tradeoffs: the language design process, support for animations within vs. between encodings, and models of animation keyframes.



\section{Related Work}
\label{sec:related-work}

Our contribution is motivated by perceptual work on the value of combining interaction and animation, and is informed by the design of existing toolkits for authoring animated data visualizations.

\subsection{Animation in Information Visualization}

In a classic 2002 paper, Tversky et al.~\cite{tversky_animation_2002} question the efficacy of animated graphics.
In reviewing nearly 100 studies comparing static and animated graphics, the authors were unable to find convincing cases where animated charts were strictly superior to static ones.
Visualization researchers have since contributed a body of studies that have identified reasons to be both optimistic and cautious about the value of animation in visualization. 
For instance, several studies have demonstrated advantages when animating chart transitions~\cite{heer_animated_2007, kim_designing_2019, dragicevic_temporal_2011, chevalier_not-so-staggering_2014} or directly animating data values to convey uncertainty~\cite{hullman_hypothetical_2015, kale_hypothetical_2019}.
However, these studies have also echoed concerns from Tversky et al. that animations are often too complex or fast to be perceived accurately\,---\,for instance, Robertson et al. found that animated trend visualizations are outperformed by static small multiples displays~\cite{robertson_effectiveness_2008}.



To ameliorate these limitations of animation, Tversky et al. suggest composing animation with interactivity, particularly through techniques that allow reinspection or focusing on subsets of depicted data. 
Robertson et al. began to probe this question by testing an interactive alternative alongside the static and animated stimuli\,---\,here, clicking an individual mark adds an overlaid line that depicts its trajectory over time.
They find that although participants are no more accurate under this interactive condition, they perform faster when using this visualization for data analysis~\cite{robertson_effectiveness_2008}.
In follow-up work, Abukhodair et al.~\cite{abukhodair_does_2013} further contextualize Robertson's results, finding that interactive animation can be effective and significantly more accurate than animation alone when users want to drill down into the data or have specific questions about points of interest. 
More recent results are similarly promising: in eye-tracking studies, Greussing et al.~\cite{greussing_learning_2020} find that interactive animated graphics not only received more attention than static or interactive-only equivalents, but these charts also produced higher knowledge acquisition in participants. 
The authors believed that the enhanced affects on memory and performance resulted from an increase in engagement and attention on the visualization, which is in line with additional research on attention~\cite{bucher_relevance_2006}. 
Our work is motivated by these results. By providing a \emph{unified} abstraction of interaction and animation, Animated Vega-Lite allows analysts to rapidly switch between the two modalities, or compose them together to best suit their needs. 
Moreover, as our abstractions preserve Vega-Lite's generative properties, we believe our contribution lowers the threshold for conducting future such studies by allowing researchers to more systematically isolate, vary, and compare individual interaction and animation techniques.

\subsection{Authoring Interaction and Animation}

In \autoref{sec:bridge}, we describe the conceptual similarities between Animated Vega-Lite and Functional Reactive Programming (FRP).
Moreover, in \autoref{sec:eval} we conduct a detailed comparison between Animated Vega-Lite and gganimate~\cite{thomas_lin_pedersen_gganimate_2019}, Data Animator~\cite{thompson_data_2021}, Gemini/Gemini\textsuperscript{2}~\cite{kim_gemini_2021, kim_gemini2_2021}, and Canis/CAST~\cite{ge_canis_2020, ge_cast_2021}.
Here, we instead survey other systems for authoring interaction and animation that have informed our approach.


Visualization toolkits such as D3~\cite{bostock_d_2011}, Plotly~\cite{noauthor_plotly_2012}, and Matplotlib~\cite{hunter_matplotlib_2007} offer a number of facilities for authoring and composing interaction and animation including typologies of techniques (e.g., brushing, hovering, and animation frames) through to event callbacks and transition functions.
Technique typologies can help foster a rapid authoring process, allowing designers to easily instantiate common techniques, but also present a sharp \textit{abstraction} cliff~\cite{thomas_rg_green_cognitive_1989}. 
If designers wish to produce more custom interaction or animation techniques, they must turn to an entirely different notation: authoring low-level, imperative event callbacks or transition functions. 
This abstraction cliff also increases the \textit{viscosity} of the authoring process~\cite{thomas_rg_green_cognitive_1989}.
For instance, to switch between the static, interactive, and animated displays of faceted data described in the introduction using D3 would involve restructuring the specification code in non-trivial ways\,---\,a problem that is exacerbated if HTML templates are used to generate the SVG rather than the \texttt{d3-selection}, as is increasingly the case when working with modern frontend frameworks such as Svelte, Vue, or React.

In contrast, Animated Vega-Lite, like its predecessor, prioritizes concise high-level declarative specification.
As \autoref{sec:vlia-by-example} describes, users can make atomic edits (i.e., changing individual keywords, or adding a localized handful of lines of specification code) to rapidly explore designs across the three modalities.
The tradeoff, however, is one of expressiveness.
Animated Vega-Lite users are limited to composing language primitives; while these primitives are sufficient to broadly cover interaction and animation taxonomies (\autoref{sec:example-gallery}), their expressive range will necessarily be smaller than their lower-level counterparts.

\section{Motivation: Unifying Interaction and Animation}
\label{sec:vlia-by-example}

In this section, we discuss similarities between interaction and animation that we observe.
These similarities drive our design decisions, allowing us to extend Vega-Lite with only minimal additional language primitives, and yielding a low-viscosity grammar that makes it easy to switch between static, interactive and animated modalities.

{
\renewcommand{\arraystretch}{1.2}

\begin{table}[t!]
    \centering
    \begin{tabularx}{\linewidth} { 
      >{\raggedright\arraybackslash\hsize=.78\hsize}X 
      >{\raggedright\arraybackslash\hsize=1.17\hsize}X 
      >{\raggedright\arraybackslash\hsize=1.05\hsize}X  }
  \toprule
     Example technique & Interaction intent \cite{yi_toward_2007} & Animation type \cite{heer_animated_2007} \\
  \midrule
    \textbf{Conditional encoding} & Select & --- \\
	\textbf{Panning} & Explore & View transformation  \\
	\textbf{Zooming} & Abstract / Elaborate &  View transformation  \\
	\textbf{Axis re-scaling} & Reconfigure & Substrate transformation  \\
	\textbf{Axis sorting}  &  Reconfigure &   Ordering  \\
	\textbf{Filtering}  & Filter &  Filtering  \\
	\textbf{Enter/exit} & Explore & Timestep  \\
	\textbf{Multi-view}  & Connect & --- \\
	\textbf{Changing encodings} & Encode &  Visualization change, Data schema change  \\
  \bottomrule
    \end{tabularx}
    \vspace{0.5mm}
  \caption{Techniques common to interaction and animation taxonomies.}
  \vspace{-5mm}
  \label{tab:taxonomies}
\end{table}

}

\subsection{Conceptually Bridging Interaction and Animation}
\label{sec:bridge}

We observe that interaction and animation share conceptual similarities at both low and high levels of abstraction.
At a low level of abstraction, Functional Reactive Programming (FRP) languages like Flapjax~\cite{meyerovich_flapjax_2009} and Fran~\cite{elliott_functional_1997}, as well as FRP-based visualization toolkits like Vega~\cite{satyanarayan_reactive_2016}, have shown that interaction and animation can both be modeled as \textit{event streams}.
The Vega example gallery demonstrates how this unified abstraction offers \textit{consistency}, with similar semantics expressed through similar syntactic forms~\cite{thomas_rg_green_cognitive_1989}.
Namely, the gallery recreates the Gapminder global health scatter plot, originally an animated visualization produced by Hans Rosling~\cite{rosling_best_2006}, but as an interactive visualization driven by the DimpVis direct manipulation technique~\cite{kondo_dimpvis_2014}.
We observe that, although it would be tedious to do manually, a user could convert this interactive visualization back to the original animated one by replacing signals near the top of the dataflow, which react to incoming drag events, with signals that respond to timer events instead: where these signals map the drag event's position to a year value, the timer signals would simply emit the next year value on each event.
The rest of the downstream reactive logic would remain unchanged.
However, as the Vega authors found~\cite{satyanarayan_declarative_2014}, additional language design is necessary to ensure FRP primitives compose together with grammar of graphics constructs and to facilitate higher-level authoring of dynamic visualizations. 

\begin{figure*}[t]
  \centering
  \includegraphics[width=\textwidth]{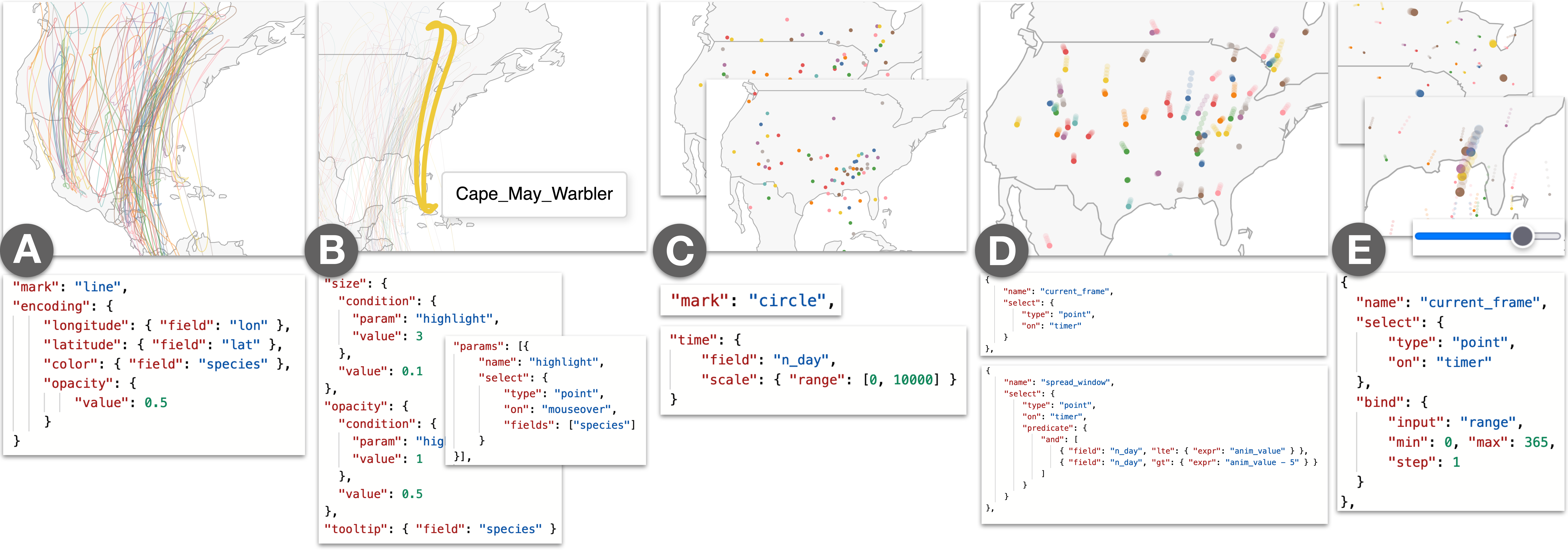}
  \caption{An analyst's workflow with Animated Vega-lite. A) Static visualization of bird migrations. B) Adding interaction to hover over a migration path and view a tooltip. C) Switching from static lines to animated circle marks. D) Adding animated path trails for the previous 5 days. E) Adding an interactive slider to scrub through the animation.}
  \label{fig:usagewalkthrough}
  \vspace{-4mm}
\end{figure*}

To analyze conceptual similarities between interaction and animation at a higher-level of abstraction, we look to \hlc[yi]{Yi et al.}~\cite{yi_toward_2007} and \hlc[heer]{Heer and Robertson}~\cite{heer_animated_2007} that taxonomize techniques for each modality respectively.
These taxonomies are defined by drawing on example visualizations, and although they have been defined separately, share many motivating techniques (\autoref{tab:taxonomies}).
For example, Heer and Robertson cite panning as an example of \hlheer{view transformation} because it changes the reader's viewpoint while leaving data schemas and encodings intact. Yi et al.\,also consider panning, categorizing it as an example of an \hlyi{explore} interaction, because it involves showing a new subset of data as points shift in and out of the viewport.
Zooming, another example of \hlheer{view transformation}, is also described as an \hlyi{abstract/elaborate} interaction because it can be used to show data at different levels of detail.
As we show in \autoref{tab:taxonomies}, we observe substantial overlap in techniques referenced by both taxonomies.
Though \hlyi{select} interactions lack an explicitly defined corresponding animation type, conditional encoding is a commonly used technique in animated visualizations.
Similarly, though there is no corresponding category in Heer and Robertson's taxonomy for \hlyi{connect} interactions, animations applied to shared backing data across multiple views can fulfill the same purpose of highlighting relationships between related points.

\subsection{Low-Viscous Authoring: An Example Usage Scenario}

A unified abstraction for static, interaction and animation also promotes a low-viscous authoring process (i.e., being able to easily switch between modalities, or compose them together).
To illustrate the affordances of this approach, we present an example walkthrough following Imani, an orthonologist, as she plans a new birdwatching expedition.
Imani has a bird migration dataset comprising the average latitudes and longitudes for a variety of bird species, for every day of the year\cite{la_sorte_convergence_2016}.
To ensure a productive trip, Imani wants uncover how migration patterns correspond to different times of the year and geographic regions.

\textbf{Static (\autoref{fig:usagewalkthrough}A).} Imani begins her analysis with a static visualization to get an overview of the dataset.
She plots a map, and visualizes migration paths using line marks: each bird species is depicted as a single, uniquely-colored line, connecting the individual daily points along their given latitudes and longitudes.
However, Imani is quickly overwhelmed as the size of the dataset produces too many overlapping lines for this static view to be useful, even after adjusting mark opacity.

\textbf{Interactive (\autoref{fig:usagewalkthrough}B).} To pick out individual bird species, and begin a cycle of generating and answering hypotheses, Imani thinks to layer some interactivity on the static display. 
She adds a \emph{point selection} named \texttt{highlight} and driven by \emph{mouseover} events. 
By default this selection is populated with the data tuple underneath the mouse cursor, and additional tuples are added or toggled when the \texttt{shift} modifier key is pressed. 
Imani writes a \textit{conditional encoding} to interactively adjust mark appearance: selected paths are drawn at full opacity and in a larger size, while unselected paths are drawn with lower opacity and at a smaller size. 
Thus, as Imani moves her mouse across the visualization, she is able to better trace individual paths, and she adds a \emph{tooltip} encoding channel to surface and note species' names.


This interactive view gives Imani a better sense of migration paths. 
But, to be able to plan her expedition, she needs to understand where different bird species may be on any given day. 
Until this point, Imani has used vanilla Vega-Lite abstractions. In the subsequent steps, we show how features of Animated Vega-Lite help Imani deepen her analysis.

\textbf{Time Encoding Channel (\autoref{fig:usagewalkthrough}C).}
Imani swaps to a circle mark and maps \texttt{day} (a field that encodes the day of the year from 0 to 365) to the new \emph{time encoding channel}.
With these two edits, each bird species is drawn as a circle indicating its location on a particular day, and the visualization animates through \texttt{day} values.
Imani can now follow the path bird species travel over the course of a year.

\textbf{Time Event Stream (\autoref{fig:usagewalkthrough}D).} Imani, however, is keenly aware that her dataset only contains average values for each species. Birds tend to appear at a given location within a small window of time around the average day in the dataset.
Thus, to ensure she does not make an erroneous conclusion, Imani wants to visualize this variability as a path trail.
To do so, she adds a new point selection named \texttt{spread\_window}, which contains a custom \emph{predicate}\,---\,a function that identifies which data tuples should be considered as falling within the selection.
In this case, Imani writes a predicate to select data from the five days previous to the current day. She does this by writing inequality expressions referencing the reserved name \texttt{anim\_value}, which stores the current data value of the animation.
In contrast to the existing \texttt{highlight} point selection, which is updated on user input events, \texttt{spread\_window} is instead populated and re-populated on every \textit{timer} tick.  
She uses \texttt{spread\_window} to dynamically filter the circle marks, ensuring only data values that lie within the selection are displayed and animated.
To visually distinguish the current day's points, she also elaborates the time encoding into an explicit selection called \texttt{current\_frame} and uses it to drive a conditional opacity encoding.
She renders current points at full opacity while rendering the trailing points at less opacity.


\begin{figure*}[t!]
  \centering
  \includegraphics[width=0.95\textwidth]{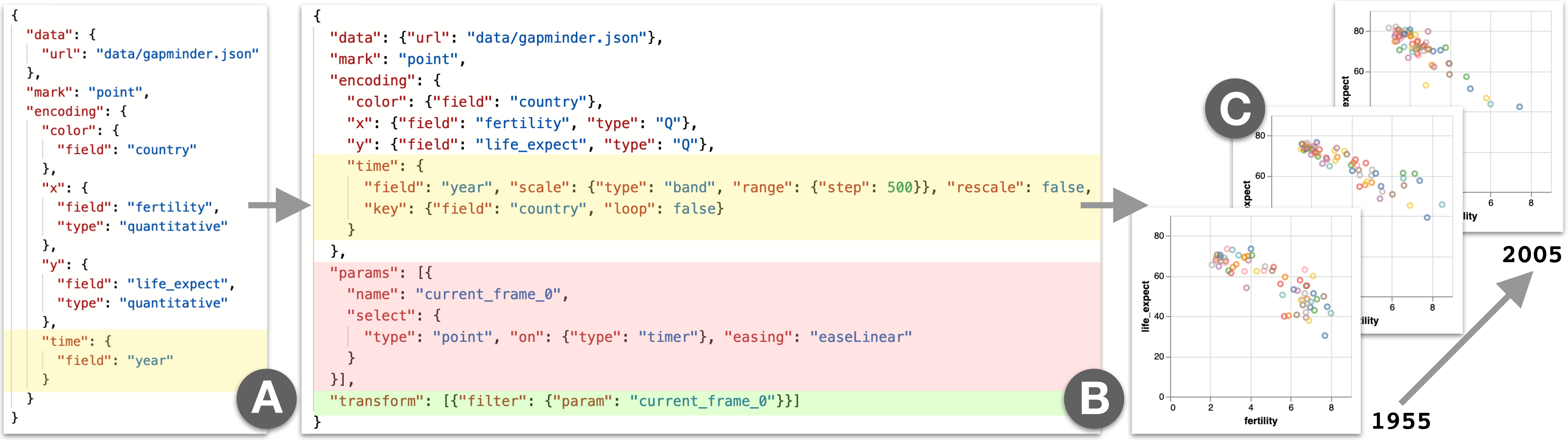}
  \caption{Animated Vega-Lite specification of the influential Gapminder animation \cite{rosling_best_2006}. (A) A minimal specification using only time encoding. (B) The same specification elaborated to show default encoding properties and a default selection. (C) Selected keyframes from the resulting animation.}
  \label{fig:gapminder}
  \vspace{-5mm}
\end{figure*}

\textbf{Composing Interaction + Animation (\autoref{fig:usagewalkthrough}E).} While watching this path-trail animation, Imani notices that a cluster of birds appear to visit Pensacola, Florida during late March and notes this region as a potential location for her expedition.
However, before she lets her colleagues know, she wants to investigate the migration patterns of the birds that come through the area\,---\,if these species tend to co-locate in other parts of the world, there is less of a reason for birders to travel to Pensacola specifically. 
To answer this question, Imani needs finer control over the animation state. 
She binds the \texttt{current\_frame} selection to an interactive range slider, and can now toggle between animating and interactively sliding the \texttt{day} field.
She scrubs the slider to the day when the birds pass through Pensacola, and to track these species in the visualization, she modifies the interactive \texttt{highlight} selection to fire on click instead of hover. 
Imani multi-selects (i.e., clicking with the \texttt{shift} key pressed) the birds that pass through the area, and then scrubs to a different day. 
Here Imani can see that these birds come from 5 unique nesting sites across the mid-west US to eastern Canada. 
This is promising as it indicates that these species uniquely overlap in Pensacola, making it a prime viewing destination.



\textbf{Summary.} With Animated Vega-Lite, Imani was able to move between static, interactive, and animated visualizations through a series of atomic edits or otherwise localized changes rather than larger-scale refactoring or restructuring of code.
Moreover, we have extended Vega-Lite's high-level affordances to animation: Imani was able to express animation as data selections and transformations, 
rather than manipulating keyframes or specifying transition states; and, the Animated Vega-Lite compiler synthesized appropriate defaults and underlying machinery for the animation to unfold correctly. 
Finally, as Animated Vega-Lite offers a unified abstraction, Imani was able to reuse Vega-Lite's existing primitives to author mixed interactive-animated visualizations as well as custom techniques without the need for special-purpose functions\,---\,e.g., combining animations with on-click highlighting and composing selections with a window data transform to draw trailing marks, rather than using a \texttt{shadow} function as with gganimate. 
\section{A Grammar of Animation in Vega-Lite}
\label{sec:grammar}

In Animated Vega-Lite, users specify animation using a \textit{time encoding channel} and \textit{timer-driven selections}.
Time encodings provide a light-weight way to convert faceted static visualizations into animations.
To further customize the animation design or easily add interaction, users can specify animations as selections instead.
Selections express dynamic data queries, and are now populated either by input events (as with vanilla Vega-Lite) or, now, via timer ticks.
Defined selections can then be used to drive data transformations, scale functions, or conditionally encode visual properties.
Our animation model expressively extends existing abstractions for static and interactive visualizations while minimally increasing language surface area and complexity.

\subsection{Time Encoding Channel}

In Vega-Lite, encodings determine how data values map to the visual properties of a mark (also known as channels).
Vega-Lite includes two channels for spatial position, \texttt{x} and \texttt{y}.
Animated Vega-Lite adds a new channel for temporal position, called \texttt{time}.
A user specifies a time encoding by providing a \texttt{field} property, which is a string of the name of a data column.
The field can be any measure type with a sort order (quantitative, temporal, ordinal), and does not necessarily need to represent a timestamp.
The system uses distinct values from this column to group data rows into temporal facets called \textit{keyframes}.
Over the duration of the animation, each keyframe is shown sequentially.

Fig.~\ref{fig:gapminder}A shows the Animated Vega-Lite specification for Rosling's Gapminder animation \cite{rosling_best_2006}.
The time encoding, highlighted in yellow, maps the dataset's \texttt{year} field to the time encoding channel.
The system uses the distinct values of \texttt{year} to group rows into keyframes. 
In other words, there is one keyframe per possible value of \texttt{year} in the dataset (i.e. \texttt{1955, 1960, 1965, ..., 2005}) (Fig.~\ref{fig:gapminder}C).

\subsubsection{Key Field}

In-betweening, more commonly called \textit{tweening}, is a standard animation technique that involves generating additional frames to smoothly transition between two keyframes.
By adding tweening, the animation will give the visual impression of continuous change over time even when data represents discrete measurements.
In data visualization, tweening takes on additional meaning as it requires generating and interpolating between values that are not present in the dataset.
In Animated Vega-Lite, to specify tweening between keyframes, the user specifies a \texttt{key} property in the time encoding channel, which references a field name.
This key field is used to group rows together across keyframes.
For two given successive keyframes, rows that share the same value for the key field are treated as the start and end states for a single mark instance.
Key values should be unique within a keyframe to prevent ambiguity; otherwise, a single mark instance might have multiple start or end states, resulting in undefined behavior.
If the user does not specify a key field, the Animated Vega-Lite compiler attempts to infer a sensible default based on the mark type and other specified categorical channels such as \texttt{color} or \texttt{detail}\,---\,an approach that follows Vega-Lite's existing inferences. 

In the Gapminder example, Fig.~\ref{fig:gapminder}B shows the Gapminder spec from Fig.~\ref{fig:gapminder}A with default values specified explicitly.
Here, \texttt{country} is used as the default key field as it is also encoded on the \texttt{color} encoding channel.
Consider the successive keyframes with \texttt{year} values 1955 and 1960.
For each year, each scatterplot point is identified by a unique \texttt{country} value.
Therefore, to tween from 1955 to 1960, the system interpolates the two rows for each country to produce the corresponding in-between point at each animation frame.

\subsubsection{Time Scale}

An encoding uses a scale function to map from the data domain to a visual range.
For spatial encoding channels, this range is measured in pixels relative to the bounding box of the rendered visualization.
For the time encoding channel, we measure the range in milliseconds elapsed from the start of the animation.
Users specify the timing of the animation using a time scale (for example, by specifying either an overall animation duration or the amount of time between keyframes as a \texttt{step}).
As with existing encoding channels, if a scale is not specified by the user, Vega-Lite infers default scale properties.
By default, scales for the time encoding channel use the unique values of the backing field as the scale domain, and create a default step range with 500ms per domain value.
For example, the Gapminder domain is a list of every fifth year between 1955 and 2005, inclusive.
The default range maps 1955 to 0ms, 1960 to 500ms, 1965 to 1000ms, and so on.
A user can override this default range to slow down or speed up the animation.

\begin{figure}
  \centering
  \includegraphics[width=\columnwidth]{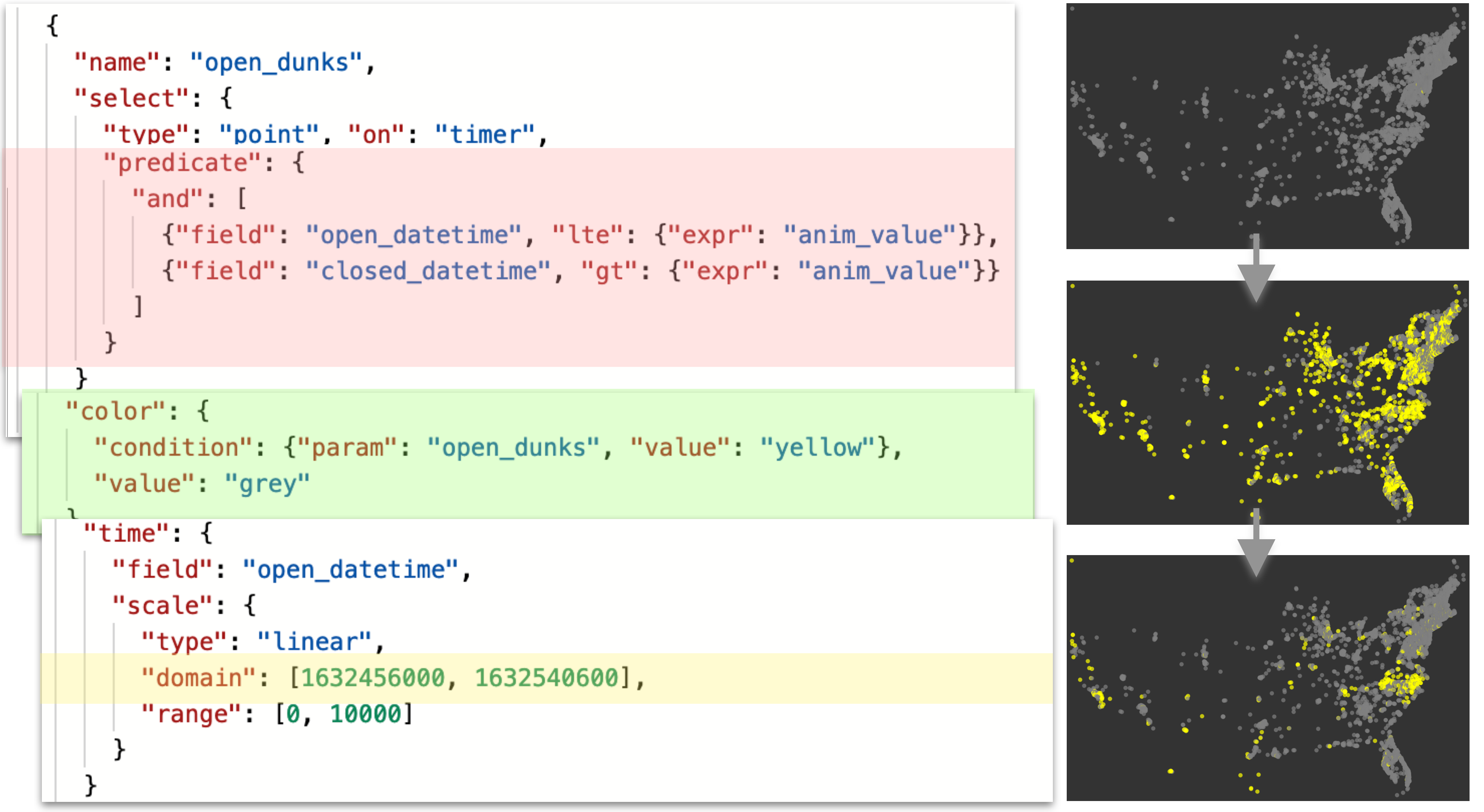}
  \caption{Animation of Dunkin' Donuts stores' opening and closing times. With a custom domain and predicate, the animation advances through 24 hours at a constant rate and conditionally colors each store if the current time is between the store's open and close times.}
  \label{fig:dunkin}
  \vspace{-5mm}
\end{figure}

Though the default domain is sufficient to express most common animations, a user may want to override the domain.
Supplying a custom domain is useful for specifying non-keyframe-based animations that require direct reference to in-between values, or require animating through values that are missing from the dataset.
For example, Fig.~\ref{fig:dunkin} shows an example of such a use case.
The animation should advance through 24-hour time span at a constant rate.
However, the dataset does not contain a field that has values that are evenly spaced in the desired domain.
So, with a default scale domain, the animation would appear to jump between time stamps rather than move through them smoothly. 
To achieve the desired behavior, the user instead specifies a custom domain representing the continuous interval between 00:00 and 23:30.



\subsubsection{Re-scale}
\label{sec:rescale}

By default, the visualization's data rectangle (or viewport) is fixed to the initial extents of the x- and y-scales calculated from the full dataset. 
However, for keyframe animations, only a subset of data is shown at any given time.
If a user wants to re-calculate the viewport bounds based on only the data included in the current keyframe, rather than the original full dataset, they can set a flag in the time encoding called \texttt{rescale}. 
When \texttt{rescale} is \texttt{true}, the viewport's bounds are recomputed at each step of the animation. 
We refer to this concept as re-scaling because re-calcuating the viewport bounds involves updating the domains of the \texttt{x} and \texttt{y} scales at each keyframe. 

\begin{figure*}[t]
  \centering
  \includegraphics[width=0.95\textwidth]{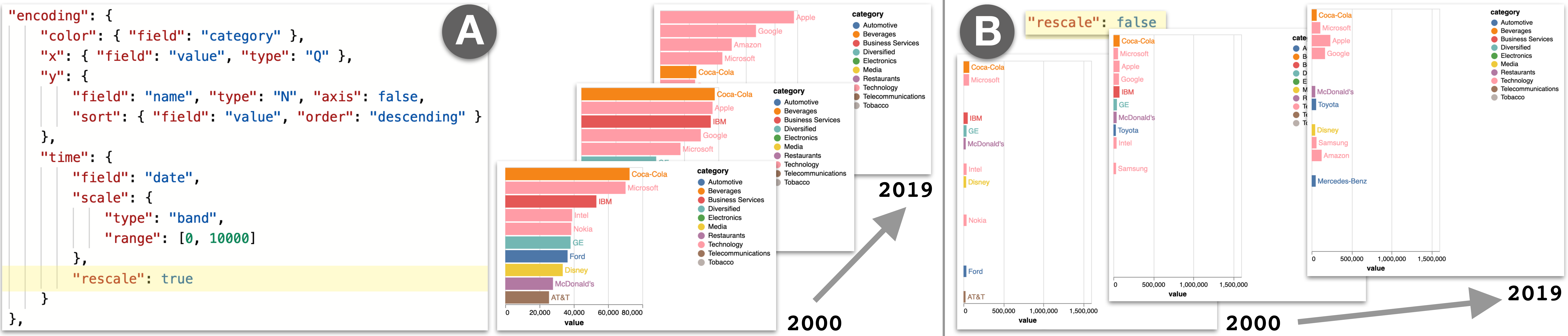}
  \caption{Demonstration of the \textit{rescale} time encoding property recreating a D3 bar chart race example~\cite{mike_bostock_bar_2019}. (A) \texttt{rescale} is \texttt{true}: the viewport is recalculated on each keyframe. (B) \texttt{rescale} is \texttt{false}: the viewport is calculated on the whole dataset, and does not update with the selection.}
  \label{fig:barchartrace}
  \vspace{-5mm}
\end{figure*}

Fig.~\ref{fig:barchartrace} demonstrates the use of \texttt{rescale}. Rescale is enabled in Fig.~\ref{fig:barchartrace}A, where the viewport updates according to the current selection. The visualization remains tightly zoomed on the currently displayed bars, with the longest bar always scaled to nearly the full width of the viewport.
In contrast, Fig.~\ref{fig:barchartrace}B has rescaling disabled. The viewport is initially calculated with the full dataset and remains fixed. This would be appropriate for Gapminder, because we want to show the countries moving along a fixed scale.
However, it is less helpful for bar chart race.
Instead of enabling positional comparisons to a fixed scale, the animation prioritizes making the ordering of the top-ranked bars salient.


\subsection{Selections with a Timer Event Stream}

\textit{Selections} are subsets of data points that are populated when updates occur in an \textit{event stream}. 
In Vega-Lite's interactive grammar, selections are defined using streams of user input events (e.g., clicks, mouse movements, or keyboard presses).
The system uses the event's properties to query a set of data points.
The selected data can then be applied to update downstream primitives in the visualization specification including data transformations, scale functions, or conditional visual encodings.
For example, a selection defined using the \texttt{mouseover} event may be used to highlight marks that a user hovers over with their cursor.
Under the hood, the selection receives a stream of \texttt{mouseover} events with \texttt{x} and \texttt{y} coordinates in pixels.
It uses the scales associated with the \texttt{x} and \texttt{y} encoding channels to invert these screen coordinates back to data coordinates (i.e. values in the domain of the corresponding scale). 
A default predicate function iterates over all rows in the dataset, and includes the rows matching those data values in the selection.

Animated selections are analogous to interactive selections. 
However, instead of reacting to input events, animated selections use a \texttt{timer} event stream to advance an internal clock representing the elapsed time of the animation in milliseconds (ms). 
This clock resets to 0ms when it reaches the end of the range defined by the time encoding's scale (i.e. the animation loops the duration of the time scale's range). 
As the clock updates, the elapsed time value is mapped to a value in the time domain (i.e. the time encoding's field values). 
The animation selection updates to include all data points matching that value.

As selections rely on scales to convert map time to data values, selection-based animations still require a time encoding channel to be defined. 
In fact, all animations that can be expressed with only a time encoding can be elaborated into selection-based animations. 
In other words, selection-based animations are strictly more expressive than animations using only time encoding.

\subsubsection{Applying Selections}
\label{sec:applying-selections}

In Vega-Lite, selections can be applied to other language constructs, including conditional mark encodings, scale domains, or data transformations \cite{zong_lyra_2021}.
This property of composition continues to hold with Animated Vega-Lite: animated and interactive selections can be used interchangeably wherever selections are supported in the Vega-Lite language.
Therefore, selections driven by timer events inherit the expressiveness of interactive selections in terms of Yi et al.'s taxonomy of interaction techniques~\cite{yi_toward_2007}.
Animations can be used to: \textit{select} marks of interest; \textit{explore} subsets of data (panning and zooming); \textit{reconfigure} data into different transformed states, \textit{connect} related items; \textit{abstract/elaborate} through overview and detail; and \textit{filter} data dynamically.
However, they cannot be used to change the properties of visual encodings on the fly, which is an interaction technique that falls outside of the selection-based model and is a limitation of base Vega-Lite.

\subsubsection{Predicate}

As the animation's elapsed time advances, the selection uses the scale defined in the time encoding to invert elapsed milliseconds (in the scale's range) to a data value (in the scale's domain). 
As a result, at any given time, there is an internal variable that has a data value corresponding to the animation's current time.
When the Vega-Lite specification is compiled into Vega, this variable is represented as a Vega signal called \texttt{anim\_value}. In the Gapminder example, \texttt{anim\_value} starts at \texttt{1955} at 0ms, and advances to \texttt{1960, 1965, ..., 2005}.

To construct keyframes, the selection queries a subset of data tuples to include in the keyframe based on the current value of \texttt{anim\_value}. 
By default, tuples are included in the keyframe if their value in the time encoding's field (e.g. \texttt{year} for Gapminder) is equal to \texttt{anim\_value}.
However, to define alternate inclusion criteria for determining keyframes, users can specify custom predicate functions. 
For example, if at every step of the animation, a user wished to show all points with year less than or equal to \texttt{anim\_value}, they would use the following predicate:

\texttt{\string{"field":\hspace{0.3em}"year",\hspace{0.3em}"lte":\hspace{0.3em}"anim\_value"\string}}

Previously, Vega-Lite did not allow users to customize the selection predicate because the majority of interactions could be expressed using a combination of default predicates and selection transformations.
Nonetheless, enabling predicate customization in the selection specification also increases the expressiveness of the interactive grammar.

\subsubsection{Input Element Binding}

Using the \texttt{bind} property, a user can populate a selection using a dynamic query widget (such as an HTML slider or checkbox).
For animated selections, input element binding offers a convenient way to add interactive playback control to the animation.
For instance, the user can bind an animated selection to a checkbox to toggle whether the animation is playing or paused.
Similarly, they can bind a selection to a range slider and drag to scrub to a specific time in the animation. 

Scrubbing the animation with the slider surfaces an interesting design challenge when combining animation and interaction: how should the system delegate control between the animation timer and user interaction?
Initially, the animation is driven by the timer, with the slider visualizing timer updates.
When the user starts dragging the slider, the system pauses the animation and delegates control to user interaction.
Pausing is necessary so that the slider does not continue to advance forward while the user is currently scrubbing.
When the user is done scrubbing, they may want to give control back to the animation.
To facilitate this, Animated Vega-Lite automatically includes a play/pause checkbox alongside bound sliders.
The user can simply re-check the box to give control over the animation back to the timer.

\subsubsection{Pausing}

Animated Vega-Lite supports pausing in two ways: by interaction, and by data value.
Interactive pauses are specified using the \texttt{filter} property of Vega-Lite event streams.
Users can provide the name of a Vega-Lite parameter to the \texttt{filter} property of a timer event stream.
Parameters can be either selections or variables.
When the provided parameter evaluates to true (i.e. is a non-empty selection or a true boolean variable), the filter will capture incoming events, preventing the animation clock from advancing.
When the paramater evaluates to false, the events will resume propagating and the animation will continue.
For example, a user can bind a checkbox to a parameter named \texttt{is\_playing}, and use the following event stream definition to pause the visualization when the box is checked:

\texttt{"on":\string{"type":\hspace{0.3em}"timer",\hspace{0.3em}"filter":\hspace{0.3em}"is\_playing"\string}}

Pausing by data value is specified using the \texttt{pause} property of an animated selection definition.
The user provides a list of data values to pause on, and the duration of each pause.
For example, a user can specify that the Gapminder animation should pause on the year 1995 for 2 seconds, to draw attention to the data for that year:

\texttt{"pause":\hspace{0.3em}[\string{"value":\hspace{0.3em}1995,\hspace{0.3em}"duration":\hspace{0.3em}2000\string}]}

\subsubsection{Global Easing}

\textit{Easing} is a common animation technique that involves controlling the rate that the animation timer advances.
Easing is typically implemented using a palette of pre-defined functions that map an animation time domain to a transformed time domain.
For example, an exponential easing function might cause the animation clock to begin advancing slowly, and then exponentially accelerate as the animation progresses.
In Animated Vega-Lite, the animation clock advances linearly by default. 
However, users can use the \texttt{easing} property of a selection to specify an easing function to apply to the whole duration of the animation. 
Animated Vega-Lite exposes D3's named easing functions~\cite{mike_bostock_d3-ease_2015}.

\section{Implementation}
\label{sec:implementation}

We implement Animated Vega-Lite using a prototype compiler, wrapping the existing Vega-Lite compiler to ingest Animated Vega-Lite specifications and output a lower-level Vega specification.
The Animated Vega-Lite prototype compiler begins by expanding a user-supplied specification into a ``normalized'' format with all implicit default values filled in explicitly.
This step includes generating default selections and transforms for animations specified using only \texttt{time} encodings, and filling in default scale and key definitions.
This normalized specification is passed to the next compiler step to simplify processing.

To convert Animated Vega-Lite into low-level Vega, we use the existing Vega-Lite compiler to make the initial conversion into Vega (using a copy of the specification with animation removed), and then call a series of functions to compile animation-specific parts of the spec and merge them with the output Vega.
Because Vega-Lite's high-level abstractions do not have a one-to-one mapping to low-level Vega concepts, seemingly-isolated Vega-Lite fragments will typically make changes in many different parts of the Vega spec.
Each of these functions takes in fragments of Animated Vega-Lite and standard Vega, and outputs a partial Vega specification that includes dataset, signal, scale, and mark definitions to merge into the output.

Compilation happens in six steps. First, \texttt{compileAnimationClock} uses definitions of animated selections and time encoding channels to create Vega signals and datasets for controlling the current state of the animation, handling pausing, and interfacing with interactive playback controls.
Next, \texttt{compileTimeScale} takes in a definition of a time encoding alongside Vega marks and scales. It creates Vega-level scales for the time encoding, and signals to handle inversions between the animation clock and the corresponding data value at that time. It also applies rescaling to mark encodings if applicable.
\texttt{compileAnimationSelections} then ingests definitions of animated selections to produce Vega signals and datasets that implement custom predicates, pausing and easing, and input element binding.
Fourth, \texttt{compileFilterTransforms} takes animation selections and any filter transforms that reference those selections, and materializes the selections as filtered datasets in Vega. These datasets provide the backing data for rendering marks at each keyframe.
\texttt{compileKey} then uses the time encoding specification to generate datasets and signals that handle tweening between keyframes.
Finally, \texttt{compileEnterExit} supports top-level enter and exit encoding definitions in Animated Vega-Lite, converting them into Vega-level enter and exit encodings.
Because of existing limitations in Vega, enter and exit currently are not well-supported for animation.
However, pending Vega support, designers should be able to control the behavior of visual encodings as marks enter and exit the current keyframe.

We chose to implement our compiler as a wrapper around the existing Vega-Lite compiler in order to facilitate rapid prototyping.
However, our current approach faces performance challenges that could be improved with internal changes to Vega and Vega-Lite.
For example, we currently support tweening by creating three separate datasets: the current keyframe, the next keyframe, and a joined dataset with tweens computed as a derived column.
This expensive operation causes noticeable lag on large datasets.
In future implementations, we can instead create a Vega dataflow operator that leverages the animation's semantics to compute tweens more efficiently.
For example, instead of computing multiple datasets independently and performing a join, the operator can create a single dataset backed by a sliding window over the time facets. 



\section{Evaluation: Example Gallery}
\label{sec:example-gallery}

\begin{figure*}[t]
  \centering
  \includegraphics[width=0.95\textwidth]{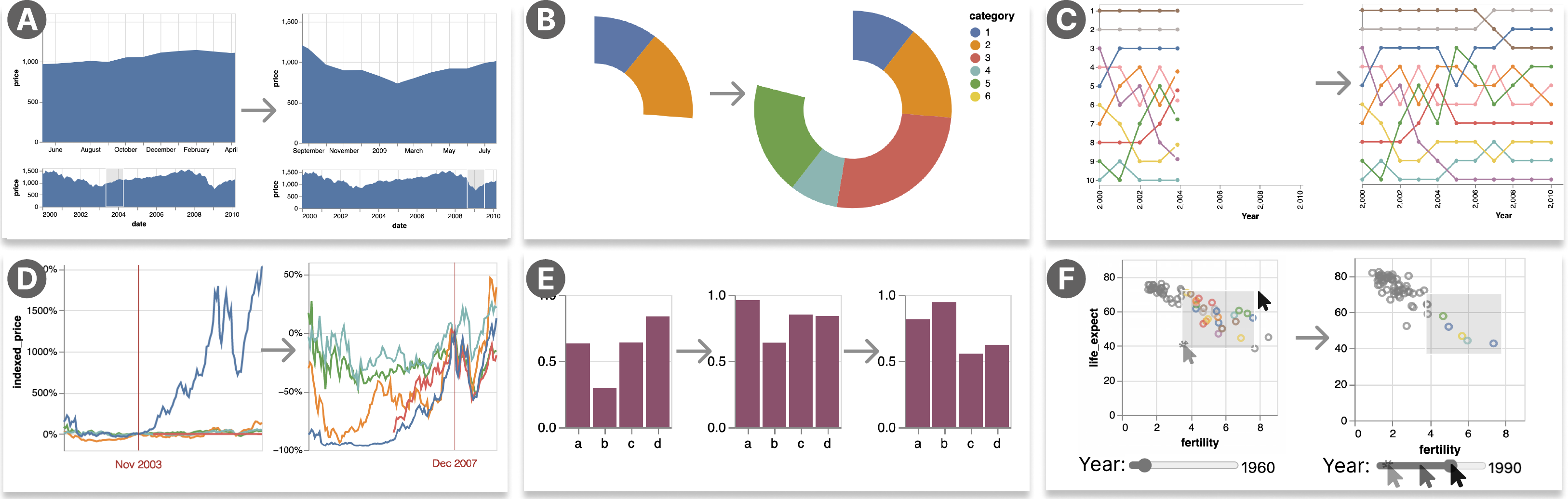}
  \vspace{-3mm}
  \caption{Animated Vega-Lite examples demonstrating coverage over interaction and animation taxonomies \cite{yi_toward_2007, heer_animated_2007} (see \autoref{fig:barchartrace} for an example \hlheer{substrate transform} and \autoref{fig:dunkin} for \hlyi{select}). A) \hlheer{View transform} via panning, \hlyi{abstract/elaborate} via overview + detail, and \hlyi{connect}ing multiple views. B) \hlyi{Filte}\hlheer{ring} data via a predicate. C) \hlheer{Ordering} / \hlyi{reconfiguring} a sorted axis in a bump chart. D) \hlyi{Exploring} sequential \hlheer{timestep}s of an index chart. E) A hypothetical outcome plot in the style of the New York Times \cite{irwin_how_2014}. F) An interactive brush selection over Gapminder.}
  \label{fig:examplegallery}
  \vspace{-5mm}
\end{figure*}




To evaluate Animated Vega-Lite's expressiveness, we created an example gallery to demonstrate coverage over both \hlc[yi]{Yi et al.'s taxonomy} of interaction intents~\cite{yi_toward_2007} and \hlc[heer]{Heer \& Robertson's taxonomy} of transition types in animated statistical graphics \cite{heer_animated_2007}.
As \autoref{fig:examplegallery} shows, we support 6\,/\,7 interaction categories and 5\,/\,7 animation categories.

\autoref{fig:examplegallery}a demonstrates an overview + detail visualization. A selection controls a brush over the bottom view, which sets the zoomed viewport of the top view. This selection is defined using a predicate that defines a sliding window over the x-axis field. When the brush is driven by animation, the selection is updated on each timer event. When the brush is driven by interaction, the selection is instead updated on drag events. Because the original Vega-Lite selection model unifies panning and zooming as selections applied to a scale domain, this approach can be adapted to animate arbitrary geometric panning and zooming behavior. This visualization demonstrates a \hlheer{view transformation}, changing the reader's viewpoint by panning and zooming the top view. It also demonstrates an \hlyi{abstract/elaborate} intent by showing the data at different levels of detail in the top and bottom view, and the \hlyi{connect} intent by showing corresponding data across multiple views.

\autoref{fig:barchartrace} shows a bar chart's x-scale dynamically recalculating on each frame using the \texttt{rescale} property of a time encoding (\autoref{sec:rescale}). This animation technique demonstrates a \hlheer{substrate transformation} through scale manipulations. It also demonstrates the \hlyi{reconfigure} intent by showing a new spatial arrangement of the data.

In \autoref{fig:dunkin} and \autoref{fig:examplegallery}b, we apply a conditional filter over the whole dataset, with filter parameters changing over time. In contrast to faceting, filtering can leverage custom selection predicates to show and hide data --- a single data point can appear in multiple groups. Both taxonomies contain a category for \hlyi{filte}\hlheer{ring}, shown here by adding or removing elements from the display. \autoref{fig:dunkin} additionally demonstrates a \hlyi{select} intent by using conditional encoding to highlight selected data.

\autoref{fig:barchartrace} and \autoref{fig:examplegallery}c show examples with a sorted axis. When a \texttt{key} is specified in a time encoding, the system automatically tweens an element's position even when its sort index has changed in the next keyframe. Continually sorting elements as the underlying data changes demonstrates an \hlheer{ordering} transition, as well as a \hlyi{reconfigure} intent.

Time encodings transition between sequential time values by default in Animated Vega-Lite (e.g. \autoref{fig:gapminder}). \autoref{fig:examplegallery}d demonstrates an additional example of this animation. A default animated point selection is applied to a data transform that re-normalizes a stock price time-series chart on each tick. The original Vega-Lite paper contains an interactive version of this example, which instead populates the point selection on mouse hover events \cite{satyanarayan_vega-lite_2017}. These examples demonstrate \hlheer{timestep} transitions, which also fulfill the \hlyi{explore} intent by showing new data points at each step. Axis re-normalization is also an example of a \hlyi{reconfigure} intent.

In addition to achieving broad coverage over the two taxonomies, our system also supports simulation techniques including hypothetical outcome plots (\autoref{fig:examplegallery}e) \cite{hullman_hypothetical_2015}.
And, as previously discussed in \autoref{sec:applying-selections}, animated selections can be applied to the same set of dynamic visual behaviors as interactive selections.
Consequently, users can easily switch between timer and input event streams when prototyping existing interaction techniques in Vega-Lite.
For example, \autoref{fig:examplegallery}a and \autoref{fig:examplegallery}d show animated selections driving common interaction techniques\,---\,panning and re-normalizing, respectively.
Users can also easily compose interaction techniques with animated visualizations by defining additional selections. For example, \autoref{fig:examplegallery}f demonstrates an interactive brush used to highlight a region of an animated Gapminder visualization. Points of interest are conditionally colored as they enter or exit the brush region.

\textbf{Discussion and Limitations.} Like the original Vega-Lite, Animated Vega-Lite intentionally trades some limits to expressivity for gains in concise, high-level, declarative specification.
In Sects.~\ref{subsec:scene-segue} \&~\ref{sec:global-local-keyframe}, we detail this expressiveness tradeoff in terms of the classes of animation techniques (Animated Vega-Lite primarily supports \textit{scene} techniques instead of \textit{segue}) as well as the implications on how keyframes are modeled and generated (Animated Vega-Lite supports non-parametric keyframe transitions, and offers some limited support for parametric keyframe transitions). 
Thus, lower-level and imperative languages will necessarily be more expressive: for instance, D3 can express both scene and segue animations, but using different language constructs (timer event loops and transition functions, respectively). 
As these sections describe, offering high-level declarative specification that unifies not only these distinct conceptual models of animation, but also interaction and static charts, remains a compelling direction for future work.

By extending Vega-Lite, Animated Vega-Lite also inherits its predecessor's limitations. 
For instance, Vega-Lite selections cannot alter visual encodings or data transformation pipelines at runtime (the \hlyi{encode} interaction type in Yi et al.'s taxonomy~\cite{yi_toward_2007}); thus, Animated Vega-Lite cannot support the \hlheer{visualization change} or \hlheer{data schema change} transition types in the Heer \& Robertson taxonomy~\cite{heer_animated_2007}.
\section{Evaluation: Critical Reflection}
\label{sec:eval}

To identify our grammar's design tradeoffs, we compared our approach to existing animated visualization grammars following the \textit{critical reflections} evaluation method~\cite{satyanarayan_critical_2019}.
We recruited five developers of existing grammars: John Thompson and Leo Zhicheng Liu\footnote{Thompson \& Liu also co-authored the original critical reflections paper~\cite{satyanarayan_critical_2019}.} of Data Animator~\cite{thompson_data_2021}, Tong Ge of Canis~\cite{ge_canis_2020} and CAST~\cite{ge_cast_2021}, Thomas Lin Pedersen of gganimate~\cite{thomas_lin_pedersen_grammar_2019}, and Younghoon Kim of Gemini~\cite{kim_gemini_2021} and Gemini\textsuperscript{2}~\cite{kim_gemini2_2021}. 
We focused on animation grammar developers because the interactive grammar was evaluated in the original Vega-Lite paper.
With each participant, we conducted a one-hour pre-interview. We then asked them to asynchronously engage with our grammar for an extended time by reading a system walkthrough and grammar documentation similar to \autoref{sec:vlia-by-example} and \autoref{sec:grammar}, respectively, and run examples similar to those found in \autoref{sec:example-gallery}. We further suggested participants write new specifications and/or port other examples, including examples from their own tools. We encouraged participants to take notes and reflect on the design of Animated Vega-Lite during the process. Finally, we conducted post-interviews with each participant that lasted 30--60 minutes. 
Each participant was offered a \$125 gift card as compensation.


Our goals were to (i) compare and contrast their design processes with ours, (ii) understand differences and design tradeoffs between their grammars and ours, and (iii) generate insights about the direction of future animation grammars. 
%
During the interviews, three of the authors of this paper began developing initial thematic hypotheses. After the interviews, we independently conducted a thematic analysis before finally coming together and synthesizing our insights, which we summarize below. These themes provide insight into the design of our grammar, and animated visualization grammars more generally.

\subsection{Grammar Design Process}

\subsubsection{Specific Examples Motivate Grammar Design}

When scoping their research projects, our interviewees prioritized motivating examples that they found personally compelling. For example, the authors of Data Animator and Gemini were both motivated in part by R2D3~\cite{stephanie_yee_visual_2015}. As we discuss in the following subsections, the choosing examples to support leads to design tradeoffs, e.g. between scene- and segue-dominant abstractions (\autoref{subsec:scene-segue}). Thus, a handful of compelling in-the-wild examples can significantly influence the grammars developers build.
Other examples that were cited across multiple interviews included Gapminder~\cite{rosling_best_2006}, Periscopic's Gun Deaths~\cite{periscopic_united_2013}, and animations in the New York Times (NYT) and the Guardian.

On the other hand, a \textit{lack} of existing examples may also motivate a grammar developer. For example, to gain more insight into the popularity of animated visualization techniques, Kim scraped NYT and Guardian articles from 2018 as well as YouTube videos from the same year. He noticed that about 90\% of the animated visualizations he studied updated data, but kept the encoding fixed. R2D3 was a notable exception. A similar imbalance can be found in the Data-Gifs example gallery~\cite{shu_what_2021}, where over half of the examples have fixed encodings. Kim hypothesized that the imbalance is influenced by the affordances of existing tools, and decided to optimize Gemini for transitions between changing encodings.

With Animated Vega-Lite, we were motivated by the large collection of existing examples with static encodings, such as those in the Data-Gifs example gallery. This category includes many prominent designs like Gapminder and bar chart races. Rather than focus on developing an expressive language of transitions between keyframes, we focused on an expressive language of keyframe generation via selections. Our abstractions facilitate the design of visualizations that must produce many keyframes backed by a fixed encoding.

\subsubsection{Natural Programming vs. Core Calculus Design}
\label{sec:natural-programming}

To make their systems easy to use for their target audiences, the authors of Data Animator and Gemini aimed to develop grammars that matched the existing mental models of animation designers. 
To that end, both groups conducted interviews prompting experienced animators to sketch interfaces or write pseudocode to recreate exemplar animated visualizations~\cite{thompson_understanding_2020,kim_gemini_2021}.
Fundamental abstractions emerged from these formative studies. 
For instance, Gemini's studies yielded the concepts of synchronizing (\textit{`at the same time'}) and concatenating (\textit{`then'}, \textit{`after'}) while Data Animator's studies surfaced designers' familiarity with keyframes in Adobe After Effects.
%
%
%
%
This design process is known as \textit{natural programming}, where a developer aims \textit{``for the language and environment to work the way that nonprogrammers expect''}~\cite{myers_natural_2004}.
%

In contrast, we set out to develop a small \textit{core calculus}~\cite{coblenz_pliers_2021} of abstractions for Animated Vega-Lite, which we outlined in \autoref{sec:grammar}. Our design was motivated by the desire to explore whether interaction and animation could be unified. This unification would likely not have been elicited by a target user.
Because the key idea of our paper is to identify a unified abstraction, this difference in approach results in a design tradeoff. As Kim explained, Animated Vega-Lite may seem natural to a Vega-Lite user, but might present a steeper learning curve to someone familiar with animation tools like Adobe AfterEffects, as Animated Vega-Lite has no explicit concept of a keyframe.

Analyzing these processes via the Cognitive Dimensions of Notation~\cite{thomas_rg_green_cognitive_1989}, we find that iterating closely with end users in a natural programming process yields a grammar that \textit{closely maps} to common user mental models. 
On the other hand, by distilling abstractions to a reduced set of orthogonal concepts, a core calculus process better emphasizes a \textit{consistent} API that has low \textit{viscosity}. 
Over-emphasizing one process or the other may drag a language design too far to one side. With PLIERS, Coblenz et al.~\cite{coblenz_pliers_2021} offer suggestions for how developers may integrate and balance between these approaches.
They recommend a developer iterate between developing the theoretical foundations of their language (core calculus) and the user-facing language (surface language). Moreover, Coblenz et al. suggest adapting natural programming by \textit{progressively prompting} a user with incrementally more information about a language's proposed API. 
This additional scaffolding can help scope how natural programming studies explore mental models, and also lets a language developer gain insights even when the core calculus significantly departs from a user's familiar models. 
Integrated design processes, like PLIERS, are likely to be valuable methods for assessing future unified grammars, because these systems must balance significant conceptual unifications with end-users' ease-of-use.









\subsection{Animation Abstractions and Design Considerations}
\subsubsection{Scene- vs. Segue-Dominant Abstractions}
\label{subsec:scene-segue}


Several interviewees noted that Animated Vega-Lite's abstractions appear complementary to their systems. For example, Kim noted his conceptual distinction between Animated Vega-Lite and Gemini is \emph{``[Animated Vega-Lite] animates the internal state within Vega-Lite, and Gemini doesn't care about the internal state. It just transforms between two static states of Vega-Lite.''} Similarly, Thompson said \emph{``if you compare [Animated Vega-Lite] directly to Data Animator, the two of them together would be really nice. What one doesn't have, the other does really well.''} 
For instance, he highlighted Animated Vega-Lite's ability to automatically generate keyframes from data (e.g., each \texttt{year} keyframe in Gapminder) and Data Animator's ability to precisely specify transitions between keyframes (such as staggering) as complementary components of the two systems. 
He also appreciated Animated Vega-Lite's ability to create overlapping keyframes via layering, as in our bar chart race example (\autoref{fig:barchartrace}).
Pedersen provides one explanation for why our approach is complementary to the existing systems we studied. In his useR! 2018 keynote, Pedersen introduced the concepts of a \textit{scene} and a \textit{segue} animation \cite{thomas_lin_pedersen_grammar_2018}. A scene animation, such as Gapminder, is one where the data is changing (such as countries ranging over years), but the visual encoding is not. One can imagine a scene playing within a fixed stage (i.e., a static visual encoding).
In contrast, a segue animation\,---\,such as a pie chart transitioning to a bar chart%
\,---\,is one where the visual encoding is changing,
but the data is fixed.
In practice, the line between a scene and segue is not always clear. For example, transitioning from a strip plot to a box and whiskers plot involves both a change to the data (computing aggregate quantities) and a change to the visual encoding (converting to box-and-whiskers).

Using this scene and segue distinction, Animated Vega-Lite and gganimate may be categorized as \textit{scene-dominant} grammars. Both systems aim to cover a large space of animated visualizations with fixed encodings, such as Gapminder and bird migrations. Both systems support an additional collection of visual encoding transformations. For example, Animated Vega-Lite supports rescaling, panning, and zooming while
gganimate supports transitions that can interpolate between different shapes with the same underlying data. Though both Animated Vega-Lite and gganimate are scene-dominant systems, Pedersen highlighted the expressiveness of Animated Vega-Lite's selection model for generating arbitrary keyframes from data (as shown with the Dunkin example in ~\autoref{fig:dunkin}) as a key conceptual distinction between the two.

On the other hand, Data Animator, Canis, and Gemini are \textit{segue-dominant}. These systems have focused primarily on connecting two distinct keyframes that may have distinct visual encodings and data. To construct a transition, Data Animator, Canis, and Gemini each construct a mapping between two keyframes. This approach works well when the data set is fixed, and there are only a few keyframes (as is typical when showing a small handful of segues). But as identified by Thompson and Liu, to support an animation like Gapminder, these systems must produce a keyframe for every year in the dataset.

As discussed in \autoref{sec:example-gallery}, Animated Vega-Lite inherits Vega-Lite's inability to represent complex runtime changes to visual encodings and data transformations. We suspect that extending Vega-Lite with these capabilities could enable segue animations in a future version of Animated Vega-Lite.
To support complex runtime changes, Vega-Lite's conditional encodings could be extended from just mark properties to mark types and data transforms as in Ivy~\cite{mcnutt_integrated_2021}.
And our support for enter and exit could be extended to operate not just on data, but also on these more expressive encoding changes.

\subsubsection{Modeling Transitions Between Keyframes}
\label{sec:global-local-keyframe}

Keyframes were the most salient animation abstraction in our interviews. We discussed keyframe concepts with every interviewee, and they would often use keyframes to pose comparisons between different systems' abstractions. Every tool had to make decisions about (i) how to generate keyframes and (ii) how to transition between them. Moreover, keyframes and transitions are useful abstractions for both scene- and segue-dominant systems. In this subsection we surface an axis of the keyframe design space: modeling transitions between keyframes.



\textbf{Non-parametric transitions.}
The simplest kind of transition between keyframes is a non-parametric transition.
Consider a linear sequence of keyframes, where each keyframe describes an entire scenegraph. 
Transitions between these keyframes are \textit{non-parametric} in that the same transition is applied to every data point. For example, changing every bar to a point in 0.5 seconds (a segue animation) is a non-parametric transition because the transition's definition is independent of the mark's encoded data\,---\,i.e. its duration is a constant value.
Similarly, animating countries in Gapminder (a scene animation) is also a non-parametric transition because the transition applied to each mark is identical (moving between two points in a fixed time interval).

Animated Vega-Lite supports non-parametric transitions via its timer, easing, and interpolation abstractions, which implicitly specify a transition across keyframes. The other libraries also support non-parametric transitions between pairs of keyframes, but only scene-dominant systems (gganimate and Animated Vega-Lite) support non-parametric transitions across \textit{many} keyframes.
In scene-dominant animations, the same transition specification can be reused across a sequence of keyframes sharing a fixed encoding.

\textbf{Parametric transitions.}
In contrast to non-parametric transitions, \textit{parametric} transitions involve transition definitions that depend on the backing data.
A common use case for this model is to stagger transitions\,---\,a common segue technique that applies a small delay to each animated element to make them easier to track~\cite{heer_animated_2007}.
Because parametric transitions depend on data, individual marks can have different timing properties during the same transition.

Segue-dominant systems Data Animator, Canis, and Gemini all support parametric transitions.
But, as Thompson identified in his post-interview, parametric transitions also increase the expressive gamut of scene animations. 
%
%
%
%
\begin{figure}
    \centering
    \includegraphics[width=0.95\columnwidth]{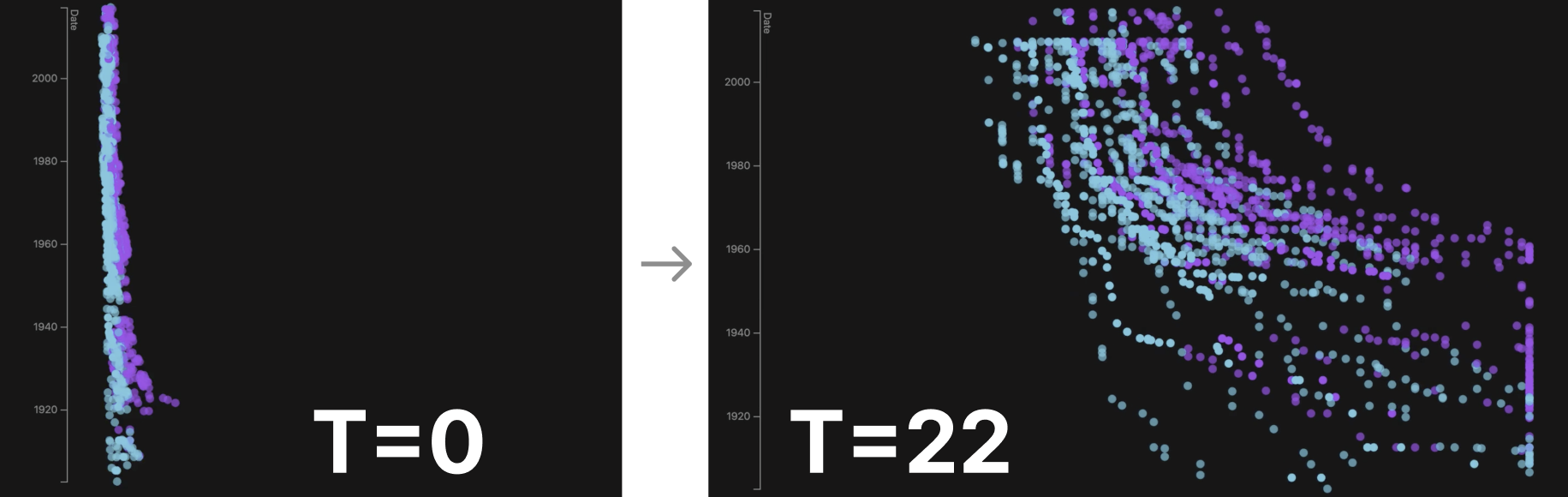}
    \caption{Swimming World Records example from Data Animator \cite{john_thompson_swimming_2020}.}
    \label{fig:swimming world records}
    \vspace{-5mm}

\end{figure}
For example, \autoref{fig:swimming world records} shows ``Swimming World Records Throughout History'' from the Data Animator example gallery. This animated scatterplot shows replays of world record swimmers. The input data includes swimmers and their final race times. When Thompson tried to port this example to Animated Vega-Lite, he realized he \emph{``had no clue how to do it. The two keyframes in this example are very simple. All of the circles at one x position, and then all of the circles like 200--400 pixels to the right. For us, you change the \emph{speed} of each individual shape based on a data property.''} 
Animated Vega-Lite could support this animation by allowing users to explicitly define a transition, with its speed parameterized by a data value.



To support parametric transitions, future versions of Animated Vega-Lite could use Lu et al.'s concept of ``dynamic functions''~\cite{lu_enhancing_2022}. These functions use mappings between data and transitions to specify rate-of-change properties of transitions over time (e.g., encoding transition speed instead of mark position).
Adapting this segue-dominant concept to Animated Vega-Lite could increase expressivity, though further work is required to understand its composition with and implications for static and interactive language constructs.
For instance, segue transition properties may more easily compose with existing static and interactive Vega-Lite constructs if translated back into scene keyframes as direct encodings instead of rates (e.g. instantiating transition speed as additional position keyframes). However, this would trade off the memory efficiency of the segue representation.




\textbf{Connecting transitions in series and parallel.}
Some of the most compelling animated examples cannot be represented as a linear sequence of transitions, parametric or not. 
For instance, Periscopic's Gun Deaths animation~\cite{periscopic_united_2013}, a visualization frequently cited by our interviewees, cannot easily be represented even by parametric transitions. 
When discussing this example, Thompson remarked:
\emph{``This was one that I had on my list of \emph{`oh it would be so cool if we could create this,'} and then I could just not figure out a way of doing it. [...] How do you have the circle appear and then drop, and then the line keeps going? I have no clue how to do that [in Data Animator]''}.
Authoring this animation is difficult because there is no linear transition specification: the animation splits in two when the circle drops and the line continues. We are not certain that \textit{any} of the grammars we have discussed in our critical reflections can easily express this animation, because it involves both scene and segue animation.

Gemini's \textit{composition rules} offer a promising path for the transitions necessary to support the Gun Deaths animation.
Gemini's \textit{concat} primitive allows a user to specify animations in series, while its \textit{sync} primitive allows a user to specify animation components that play in parallel. Using these primitives, one could specify a sync that splits the animation into the circle and the line, and then concat the many stages of the Gun Deaths animation together. More generally, concat and sync allow a user to model transitions as a \textit{series-parallel} graph~\cite{wikipedia_contributors_seriesparallel_2022}.

However, this abstraction alone is not enough. While Gemini has a rich transition language, it cannot generate keyframes automatically from data like Animated Vega-Lite. This generation is necessary for the Gun Deaths animation to visualize individual points. Combining Gemini's segue abstractions with Animated Vega-Lite's scene abstractions is a promising future direction for expressive animation.

\section{Conclusion and Future Work}
\label{sec:discussion}

Animated Vega-Lite contributes a low viscosity, compositional, and systematically enumerable grammar that unifies specification of static, interactive, and animated visualizations. 
Within a single grammar, authors can now easily switch between the three modalities during rapid prototyping, and also compose them together to effectively communicate and analyze faceted and time-varying data.

Our grammar takes a promising step in helping authors develop visualizations that leverage the dynamic affordances of computational media.
During interviews, Pedersen described unification as the ``holy grail'' of data visualization APIs: \emph{``A grammar of graphics that defines how things look, a grammar of animation that defines how things react, and a grammar of interaction that defines how things interact. Having all of that in one unified theoretical framework would simply be awesome.''}
Future work might more deeply explore the distinctions and tradeoffs we surfaced between transition and keyframe models, or study the implications of unification at the lower-level of reactive programming semantics and data stream management.

Beyond language design, we hope that Animated Vega-Lite facilitates future work on interactive and animated visualization akin to the role the original Vega-Lite has played.
For instance, how might we leverage Animated Vega-Lite's ability to enumerate static, interactive, and animated visualizations to study how these modalities facilitate data analysis and communication\,---\,replicating and extending prior work~\cite{robertson_effectiveness_2008} more systematically? 
Similarly, how might study results be codified in the Draco knowledge base~\cite{moritz_formalizing_2019}, or exposed in systems like Voyager~\cite{wongsuphasawat_voyager_2016, wongsuphasawat_voyager_2017} or Lux~\cite{lee_lux_2021} to recommend animated visualizations during exploratory data analysis? To support this future research, we intend to contribute our work back to the open source Vega-Lite project. 

\acknowledgments{
We thank our critical reflections interlocutors and anonymous reviewers. This work was supported by NSF grants \#1942659 and \#1900991 and by the NSF's SaTC Program. This material is based upon work supported by the National Science Foundation under Grant No. 1745302.}

\bibliographystyle{abbrv-doi}

\bibliography{interaction-animation}
\end{document}